# A New Efficient Key Management Protocol for Wireless Sensor and Actor Networks

Yunho Lee , Soojin Lee

Department of Computer & Information Science,
Korea National Defense University,
Seoul, South Korea

*Abstract*—Research on sensor networks has become much more active and is currently being applied to many different fields. However since sensor networks are limited to only collecting and reporting information regarding a certain event, and requires human intervention with that given information, it is often difficult to react to an event or situation immediately and proactively. To overcome this kind of limitation, Wireless Sensor and Actor Networks (WSANs) with immediate-response actor nodes have been proposed which adds greater mobility and activity to the existing sensor networks. Although WSANs share many common grounds with sensor networks, it is difficult to apply existing security technologies due to the fact that WSANs contain actor nodes that are resource-independent and mobile. Therefore, this research seeks to demonstrate ways to provide security, integrity, and authentication services for WSAN's secure operation, by separating networks into hierarchical structure by each node's abilities and provides different encryption key-based secure protocols for each level of hierarchy: Pair-wise key, node key, and region key for sensor levels, and public key for actor

*Keywords ; Wireless Sensor and Actor Network(WSAN), Key management Protocol*

## I. Introduction

Sensor networks, which have gained interests with the advancement of wireless communication technologies and embedded computing, are being widely adapted into many applications and many active researches on related subject are being carried out. A sensor network utilizes multitudes of sensor nodes within or neighboring the area of event to collect integrate, process, and relay the information regarding the event through sink node. Due to this inherent structure, this system requires additional special efforts in order to enable immediate and on-time response to the events based on those processed information.

WSANs(Wireless Sensor and Actor Networks) are proposed to overcome the limitations of traditional sensor networks. It includes mobile and resource-efficient actors within the network and enables these actors to respond appropriately based on the information collected by sensor nodes. This is in fact a very useful and applicable type of network that can be used in applications such as Forrest-fire monitoring, home intrusion prevention or military surveillance and operations [1].

WSANs share many similarities with sensor networks as they are networks without infrastructure and they use wireless communication technologies. Therefore WSANs require many existing applied technologies in their deployment. Unlike traditional sensor networks whose nodes share the same authority and power, actor-based WSANs require a different approach in implementing these technologies. Especially, WSN only consists of sensor nodes which are resource dependent. And the network structure of WSNs is very simple. Considering both the resource limitation of sensor nodes and the structural simplicity of WSNs, most key management protocols have researched by symmetric encryption approach. But WSANs have not only sensor nodes but also actor nodes which are resource independent. Thus, the structural feature of WSANs has to be considered. Due to the facts, existing protocols for the WSNs are not suitable for WSANs. Therefore, in this paper, we propose a new efficient key management protocol for the WSAN. The major contributions are summarized as follows:

  a) Our proposed protocol splits the WSAN into two layers, the upper (sink-actor) layer and the lower (actor-sensor) layer. In the lower layer, we use symmetric approach by adopting the key management concept of the LEAP. But, especially to achieve the energy efficiency, we reduce the number of the key and simplify the procedure of the key establishment, and then reduce the amount of memory required and communication cost comparing with LEAP. In the upper layer, we use asymmetric encryption mechanism to provide the high degree of security.

  b) When replacing the existing actor node with the new actor node, it is not efficient that again establish the node key and region key between the new actor node and sensor nodes. So we employ the binding table in the actor and sink node.

This paper examines the communication structure and security requirement for WSANs, and proposes a new efficient key-management protocol to ensure security in routing, transmission, and authentication of data. Performance analysis of the proposed protocol will follow and demonstrate its security and efficiency.

The remainder of this paper is organized as follows. Section 2 summarizes previous researches. Section 3 draws the security requirements through network structure analysis.

Section 4 proposes an efficient key-management protocol for WSANs. Section 5 analyzes the security of our scheme. Section 6 analyzes the performance of proposed protocol. Finally, section 7 will find a summary and conclusion of all above sections.

## II. RELATED WORKS

The objective of security in a sensor network is to ensure confidentiality, authentication, integrity, and availability using the existing network capabilities. To achieve this objective the researches on sensor network security have been occurring in three major branches: first the sensor network security service structure approach, offers authentication through a Trustee relationship suitable for sensor networks [2][3][4]; second the key management approach based on a random subset key pre-distribution from a large key pool[5][6][7][8][9][10][11]; third asymmetric cryptography approach[12][13].

Perrig et al. proposed SPINS, a security architecture specifically designed for sensor networks [2]. The structures of SPINS are comprised of SNEP (Secure Network Encryption Protocol) which offers data security, authentication, and resetting keys to prevent repeated attacks. The u-TESLA Scheme provides authentication for broadcasted data. This method requires all sensor nodes to pass through the base station for security keys, resulting in heavy traffic overhead and extended delays when there are too many nodes to authenticate. It also requires all nodes to synchronize its time to work properly. S. Zhu et al. proposed LEAP (Localized Encryption and Authentication Protocol) which can overcome eavesdropping of data and limitations on resources and computing power of sensor nodes through encryption and source authentication [3]. Unlike previous single-key methods, LEAP uses 4 different types of keys that are used for each different type of messages being transmitted. The four keys are: Individual key shared by Base Station(BS) and all nodes, Pair-wise key shared by one neighboring node within one hop of a given node, Cluster key shared by all neighboring nodes within one hope of a given node, Group key shared by everyone in the network. Individual key and the group key are pre-saved before nodes are deployed, and u-TESLA scheme renews the group key within the predefined intervals. Assuming that the base station is safe, there is no need to consider the safety of this group key. The pair-wise key is generated based on the initial key. The cluster key is encrypted by this pair-wise key before getting transmitted.

Eschenauer and Gligor proposed a random key pre-distribution scheme [5]: before deployment, each sensor node receives a random subset of keys from a large key pool. Based on the [5], Chan, Perring, and Song proposed a *q*-composite random key pre-distribution scheme [6]. The difference between this scheme and the Eschenauer-Gligor scheme is that q common keys, instead of just a single one, are needed to establish secure communications between a pair of node. It is shown that, by increasing the value of q, network resilience against node capture is improved, i.e., an attacker has to compromise many more nodes to achieve a high probability of compromised communication. Du et al. proposed a new key pre-distribution scheme [7], which significantly improves the resilience of the network compared to the existing schemes. This scheme exhibits a nice threshold property: when the number of compromised nodes is less than the threshold, the probability that any nodes other than these compromised nodes are affected is close to zero. A random key pre-distribution scheme that uses deployment knowledge was proposed by Du et al.[8] and Huang et al.[9] and Lee et al.[10]. Dai et al. recently proposed a new key pre-distribution scheme based on Rooted- Tree in WSAN [11]. The key management tree is constructed where sink is the root, actors are the branches and sensors are the leaves, to achieve the distributed and integrated key management. One drawback of this key management approach is that some wireless links may not be keyed and thus a node may need to use a multi-hop path to communicate with one of its neighbor nodes. Since each sensor node should generate and then store many keys to share with all its neighbors immediately after deployed, the communication and storage cost are generally huge.

Several other methods based on asymmetric cryptography are also proposed: Zhou and Hass proposed a secure ad hoc network using secret sharing and threshold cryptography [12]. Kong et al. also propose localized public-key infrastructure mechanisms, based on secret sharing schemes [13].

Usually, asymmetric cryptography mechanisms ensure a powerful security in authentication. However, this mechanism requires more cost to authenticate between nodes, thus is suitable for nodes which have enough resource.

## III. NETWORK STRUCTURE AND SECURITY REQUIREMENTS

### A. Network structure

The basic structure of WSANs is shown in figure 1. The main different between WSANs and an existing sensor network is that there are actor nodes in between sensor nodes and sink node. These actor nodes have larger capacity, more computing power, better communications ability, and stronger mobility.

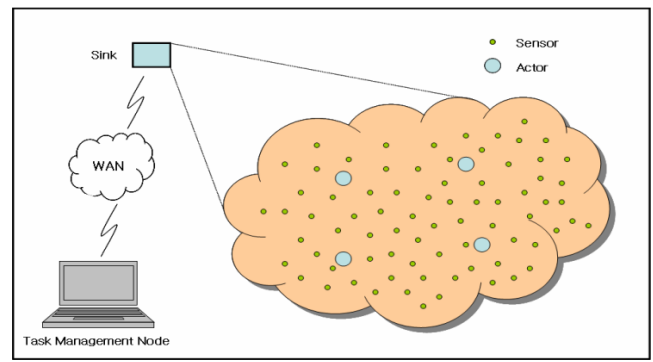

<Figure 1. Network structure of WSANs >

Therefore WSANs plan and operate the network based on the different capabilities of the three different types of nodes within the network. Sensor networks are located in a designated location and respond to the requests of actor nodes or sink node. Actor nodes send out queries to sensor nodes in the neighboring area then integrate the responses of sensors nodes and take the appropriate action when necessary. Finally, sink node work similar to actor nodes to control activities of sensors or actors when required. Sink node also plays a role of a gateway to an external network, allowing external task manager nodes to monitor and control the entire network.

Communication between actors is possible either in single or multiple hops. For actor-sensor communications, actor can communicate in one hop with a sensor node within its area of responsibility while a sensor node requires setting up a multiple hop traffic path to an actor. Figure 3 shows the conceptual diagram of network communication in WSANs.

Since actors in WSANs are mobile, the region each actor is responsible for may change at any time. When this occurs, actors can coordinate with each other to redistribute the area of responsibility.

### B. Security requirements

By the nature of wireless communication, any data packets in WSANs may be exposed to attackers. Attackers may choose to threaten the confidentiality and integrity of packets using various methods and may even carry out attacks like DoS in order to destabilize the network.

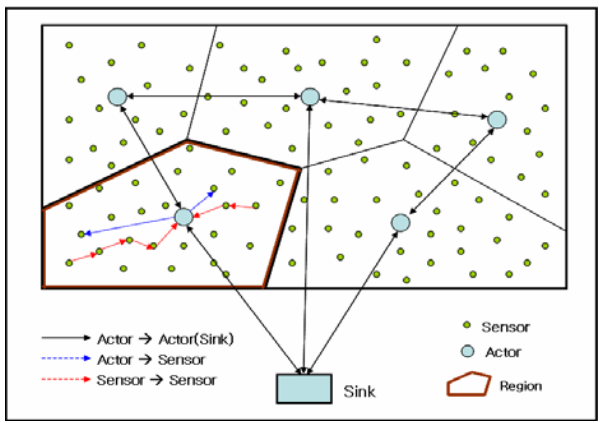

<Figure 2. Communication concept of WSANs>

Exposure of sensitive information such as encryption keys and their destination to malicious attackers may threaten the entire network. Therefore, sensitive information must be kept secure using encryption. Packets must also be authenticated in order to prevent malicious codes from being injected into packets that are transmitted. A receiver must be able to determine whether the sender is a trustworthy source and also whether the original message sent by the sender is kept intact during its travel. Freshness of data must be kept as well in case of repeated attacks. Therefore, confidentiality, authentication, integrity, and availability must be provided for all data transmitted within the network.

### IV. A KEY MANAGEMENT PROTOCOL FOR WSAN

#### A. Assumptions

First, sensor nodes are limited in their storage capacity, computing power, and communication ability. They are distributed randomly and once deployed their positions are fixed. Actor nodes may move freely within their region of responsibility and can communicate within one hop to all nodes located in the region. They have enough storage capacity and computing power.

Second, due to the weakness of wireless communication, all data packets may be exposed to an attacker. The proposed scheme has the powerful intrusion detection system to detect any malicious attacks from an attacker. Also we assume that both actor nodes and sink node will not be compromised.

Third, all nodes are given an initial key and a certificate from a third party in order to create pair-wise keys and node keys. They also contain a very strong encryption algorithm.

Finally, the time $T_{exposure}$ which is the duration of exposure of the key information of a node taken over by an attacker at the time of initial nodes distribution is always longer than the time $T_{discovery}$ which is the duration of detection and key generation of neighboring node by a node.

#### B. Overall setput and notations

The approach this paper uses is based on the differentiated network structure of WSANs as shown in figure 3. Since the layer higher than actors has enough resources and computing power, we proposed an approach based on asymmetric encryption mechanism. For the layer lower than actors which has limitations in capacities and power, we proposed an approach using symmetric encryption. Each sensor and actor node has a unique id and maintains three key types based on LEAP scheme.

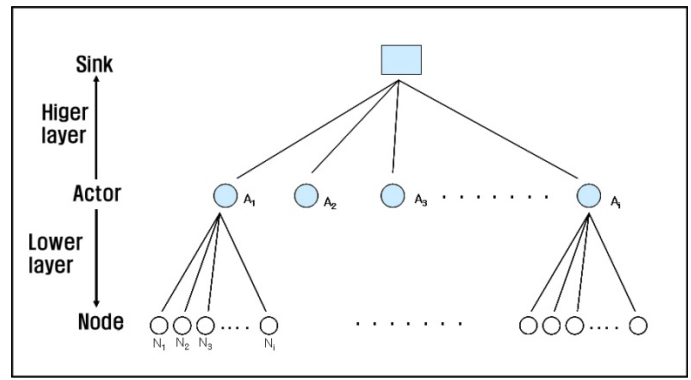

<Figure 3. Hierarchical structure of WSANs>

For lower layer below actors, the following key management structure comprised of three keys ensures security of messages transmitted from actors and reports produced by sensor nodes.

**Pair-wise key:** This key is shared by sensor nodes within one hop neighbors, and used to ensure integrity of data when a sensor node responds to a request by an actor. The actor is considered as a sensor when generating this key.

**Node key:** This key is shared by each sensor node in a region with actor, and used for encrypting sensitive information being broadcasted to the network such as reports or key renewals.

**Region key:** This key is shared by sensor nodes within a region of responsibility of an actor, and used to encrypt and authenticate messages or commands broadcasted to sensors within a specific region managed by the actor.

For the upper layer the security requirements are met using general public key structure appropriate for MANET. This public key system is used to ensure security between an actor and sinks. It includes private key, public key, and certification.

Table 1 summarizes notations used in this paper.

< Table 1. Notations>

| Notation | Description |
|---|---|
| $S(Sink)$ | Sink node or Base Station |
| $A$ | Actor node |
| $PK_A, SK_A, CERT_A$ | Public key, Secret key and Certification of actor A, respectively |
| $PK_{CA}, SK_{CA}$ | Public and Secret key of Certification Authority |
| $S_1, S_2 \ldots S_i$ | Sensor nodes |
| $M_1 \| M_2$ | Append message $M_2$ to message $M_1$ |
| $K_{IP}, K_{IN}$ | Initial keys for pair-wise and node key generation |
| $E_k[msg]$ | Encrypt *msg* using the key *k* |
| $D_k[msg]$ | Decrypt *msg* using the key *k* |
| $f()$ | Unidirectional hash function |
| $MAC(key,msg)$ | Message authentication code generated by the key |
| $F_K$ | Pseudo-random number Function |
| $N_{S1}$ | Random number generated by node $S_1$ |
| $K_{Si}$ | Master key for node $S_i$ |
| $K_{S1S2}$ | Pair-wise key for nodes $S_1$ and $S_2$ |
| $K_{Ni}$ | Node key between actor A and node $S_i$ |
| $K_{Rid}$ | Region key between actor A and regional sensor nodes $S_i$ |

## C. Key management for lower layer

### 1) Establishing Pair-wise Keys

In this scheme, the pair-wise key is targeted to the entire network and does not differentiate the type of nodes when establishing the key.

**Pre-distribution phase:** The initial key $K_{IP}$ is generated and saved into memory. Each node then produces own master key using the pseudo-random number function out of the initial key. For example, the master key for the node $S_i$ is as follows:

$$K_{Si} = F_{K_{IP}} (ID_{Si})$$

**Discovery phase:** After nodes are deployed, all nodes broadcast their ID and random number. Upon receiving this broadcast, neighboring nodes generate MAC using their own master key and send it back to the original node. Sources are authenticated by verifying the MAC received from neighboring nodes.

**Pair-wise key generation phase:** After verifying the source, each pair of nodes generate same pair-wise key using pseudo-random number function on the other's ID. Pair-wise key for node S1 and S2 are generated as follows:

$$K_{S1S2} = F_{KS2} (ID_{S1})$$

**Key deletion phase:** Sensor nodes which finish the initial key setup completely delete all keys used in the process from the memory, except their own master key.

Pair-wise key establishment is illustrated in figure 4.

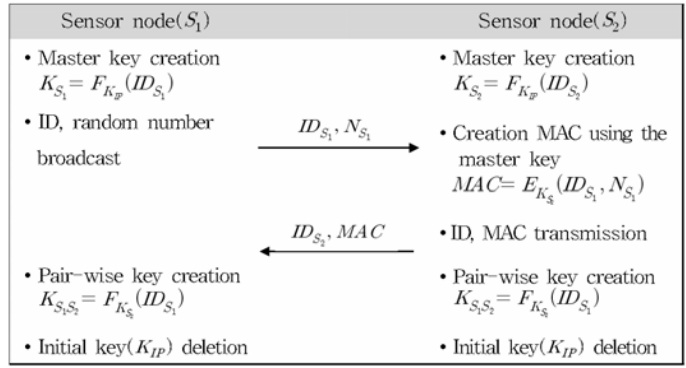

<Figure 4. Pair-wise key establishment>

### 2) Establishing Node Key

An actor broadcasts its presence and its ID to the region. Sensor nodes encrypt their own IDs using the previously distributed initial key ($K_{IN}$) and send it to the actor. The actor node then generates the node key using the ID of the sensor and randomized number, and sends it back to sensor nodes. After establishment, sensor nodes delete previously distributed key from its memory. The following figure 5 is the summary of the node key establishment procedure between a sensor node and an actor node.

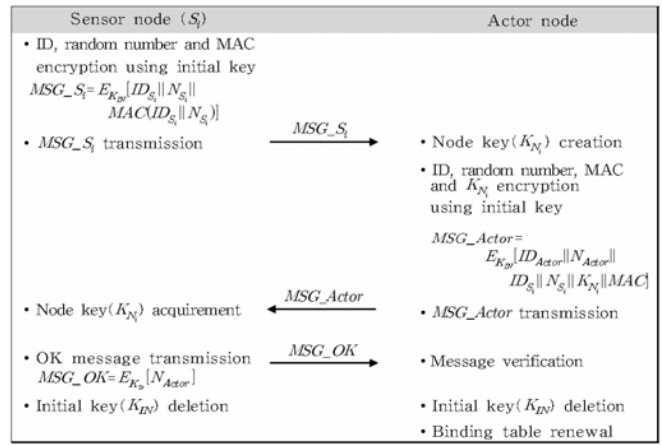

<Figure 5. Node key establishment>

### 3) Establishing Binding Table

After establishing node key, an actor node keeps a binding table which saves all sensor node IDs within the region and their node keys. Actor nodes also keep their own binding tables on the binding tables of sink node. Figure 6 shows a sample binding table.

| Number | Node_ID | Key_Info |
|--------|---------|----------|
| 001 | 0x B42DA56E | 0x cd4f12a3 |
| 002 | 0x 49EB19F8 | 0x 8b49d71a |
| 003 | 0x 7D3B4821 | 0x 20b47a3f |
| . | . | . |
| . | . | . |
| . | . | . |

<Figure 6. Example binding table>

This not only allows convenient update for the binding table establishment of a new actor, as using the binding table of sink node without going through whole key generation procedures, but also relives energy consumption for energy-constrained sensor nodes.

*4) Establishing Region Key*

Actor nodes generate a random key $K_{Rid}$ using random numbers. $R_{id}$ signifies the identification of the region which the actor is responsible for. This region key is encrypted using node key $K_{Ni}$ and distributed to each sensor node. The following figure 7 is the summary of region key establishment procedure between a sensor node and an actor node.

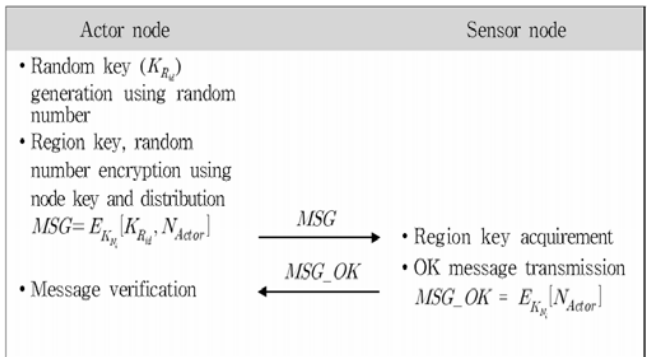

<Figure 7. Region key establishment >

The actor sets an expiration time during the region key generation, and when the key expires the actor broadcasts notifications to all sensor nodes within the region and redistributes a new region key.

*5) Adding Nodes*

The addition of new sensor nodes must be easy and fast in order to create large-scale WSANs. Additional sensor nodes are given initial keys of $K_{IP}$ and $K_{IN}$ before being deployed to generate pair-wise and node key, respectively. When a new sensor node is being added to the region, The task manager node must retransmit the initial key to the actor node for establishing the node key.

Sensor nodes establish the node key using above-mentioned processes and delete the initial key when it is done. Then the region key is established when the actor notifies the timeouts and renewals of region key.

*6) Nodes Deletion and Key Renewal*

In the WSANs, since sensor nodes are mainly used in hostile environment, their function may be lost. If the sensor node is compromised by a malicious node, its keys can be exposed. This situation causes serious problem in the WSANs. So when the sensor node is compromised and then excluded in the network by any detection algorithm, we have to ensure the network security by renewing the keys. The proposed procedures in this paper are done in following steps.

**Phase A** – Detection: Based on intrusion detection algorithm for wireless sensor networks detect and recognize the attack or loss of node by a malicious attacker. WSANs can use this to detect any anomalies within network and start to find a solution for it.

**Phase B** – Removing pair-wise key: All nodes within one hop of the affected node remove their pair-wise key. Attacker is no longer able to use that key to continue further attacks to other parts of network.

**Phase C** – Renewing region key: Actor node broadcasts the region key renewal message to all nodes within the region. The original region key no longer becomes valid and prevents the lost node from disguising itself as the actor node. The real actor node then removes the node key of lost node and updates the binding table.

**Phase D** – Reset region key: Actor node recreates the region key and transmits the key to all effective nodes within the region by unicast, excluding the lost node.

Figure 8 depicts the above procedure.

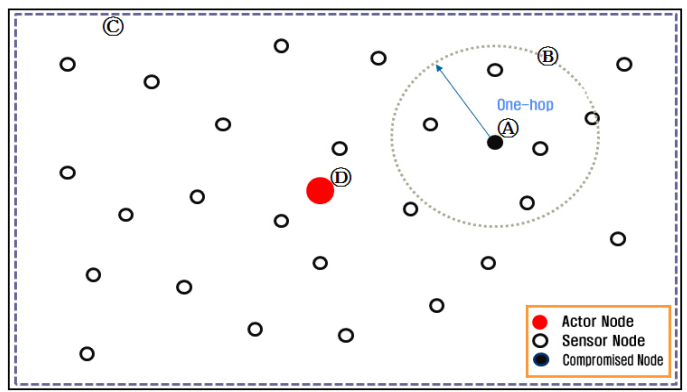

<Figure 8. Procedure of key renewal>

*D. Key management for upper layer*

The upper layer utilizes publicly available public key encryption method applied to MANET. In this proposed method the sink node plays a role of a Certification Authority. Key management and authentications are done in following steps:

**Pre-distribution of keys and certificates:** Actor node A receives $PK_A$, $SK_A$, $CERT_A$ and $PK_{CA}$ from a trustworthy third-party on offline source. In this process, the following equation is produced:

$$CERT_A = E_{SKCA}[\ A \parallel PK_A \parallel T_{sign} \parallel T_{expire}]$$

$T_{sign}$ is issuance time and $T_{expire}$ is expiry time for the certificate. Each certificate has same expiry period. Sink node saves $PK_{CA}$, $SK_{CA}$, and $CERT_{CA}$ on the binding table, as well as each actor's ID and their public keys.

**Destroying certificates:** When a node becomes captured by an attacker or loses its function, the certificate for that node becomes no longer valid and it is saved onto Certificate Revocation List (CRL) for records.

**Certificate authentication and session key generation:** Two nodes in the upper layer (actor-actor or actor-sink) follow an authentication procedure as shown in figure 9.

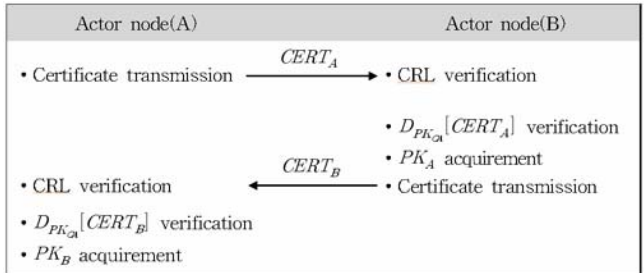

<Figure 9. Certificate authentication >

After accomplishing the authentication procedure, each node obtains the public key of the other. Using this public key, the session key for the pair node is produced by following procedure as shown in figure 10 and used for one communication period. If the communication is complete, the session key is immediately deleted.

**Certificate renewal:** The certificate renewal uses the predefined constant $T_{refresh}$. The renewal time, expiry time, and $T_{refresh}$ have the following relationship.

$$T_{refresh} \leq (T_{expire} - T_{sign})$$

All certificate owners must renew their certificate within time $T_{refresh}$.

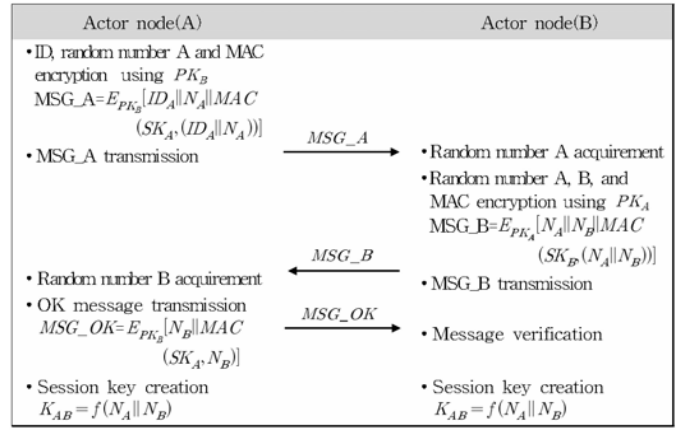

<Figure 10. Session key establishment >

## V. SEcurity analysis

In this section, we analyze the security of our scheme. As mentioned in Sec. 4.1, both actor nodes and sink node are will not be compromised. So we only analyze the security of our scheme in the lower layer. The proposed protocol in this paper is an ID-based scheme which authenticates all messages traversing; therefore an internal attacker is the only point to be considered. In case the compromised node modifies or cast away the packet, the next sensor node can detect this by overhearing the message sent out of the node. Most of the attacks fall into one of following categories: spoofed, altered, or relayed routing information, sinkhole, wormhole, Sybil, HELLO flood attacks. We discuss how our scheme can defend against those.

- Spoofing, altering, or replaying routing information: We assume that our scheme has the powerful intrusion detection system. Once the compromised node is detected somehow, our nodes deletion and key renewal scheme [see section 4.3.6] can efficiently eliminate that node from joining routing process.

- Wormhole and sinkhole attacks: An insider adversary simultaneously needs to compromise at least two sensor nodes to create a wormhole. But our scheme can enough easily detect the compromised node by using the intrusion detection system. Thus it is difficult for the adversary to create a wormhole without being detected. In a sinkhole attack, a compromised node may try to attract packets from its neighbors and then drop them, by advertising information such as high remaining energy. But neighbor nodes can detect this by overhearing the node. If malicious actions are detected, our nodes deletion and key renewal scheme can efficiently defend.

- Sybil attacks: In the key establishment phase, MAC and random number are used to authenticate between the sender and the receiver. Therefore, a node cannot pretend to be another node. As a consequence, Sybil attack would not work.

- HELLO flood attacks: The attacker may try to launch a HELLO Flood Attack in which it sends a HELLO message to all the nodes. However, this attack will not succeed in our scheme because every sensor node only accepts packets from its authenticated neighbors. So our scheme can prevent HELLO flood attacks.

## VI. PERFORMANCE EVALUATION

We evaluate the performance of our scheme by comparing with LEAP. The evaluation was carried out mathematical analysis. This paper ignores the computation load, storage requirements or communications cost for actor nodes and sink node since they already have enough capacity. Thus, considering only the sensor nodes, which has limited amount of capabilities, same performance analysis as LEAP can be applied to determine the communication and computational cost. Sensor nodes generally communicate with neighbors within one hop to generate and transmit keys. The costs are thus determined by the density of nodes, not the size of network.

The computational cost in the proposed method can be obtained by the number of messages get encrypted or decrypted. That is, a sensor node performs one hash computation to create a master key from a initial key, and costs are incurred when encrypting messages using the master key and also while generating pair-wise keys. In a network of size N where every node has a connection degree d, the computational cost can be obtained by the following equation.

$$(2d+1)N + 4N + 2N \text{ (Eq. 1)}$$

In Equation 1, $(2d+1)N$ is the computational cost for generating pair-wise keys of the entire nodes. $4N$ is the cost for node keys and $2N$ is for the region keys. If network size is fixed, the computational cost is dependent on the density of nodes and complexity becomes $O(d)$. When comparing the proposed method to LEAP [3] in computational costs, it can be demonstrated that this method is more efficient since the complexity of LEAP becomes $O(d^2)$ when network size is fixed like above assumption. Figure 11 shows the graphical representation of computation complexity of LEAP and proposed method.

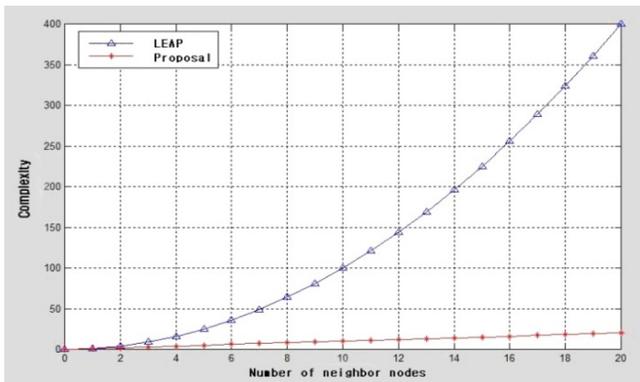

<Figure 11. Comparison of computation complexity>

The communication cost is the sum of all costs incurred during transmission and reception of the messages while each node generates appropriate keys such as pair-wise keys, node keys, and region key. While generating pair-wise keys, each sensor nodes incur one unit of communication cost to broadcast its existence to neighbors, and another unit is spent by the neighboring node for response. Node key requires 3 units, and region key requires 2 units. Therefore, the total sum of cost in this proposed protocol is as follows:

$$2dN + 3N + 2N \text{ (Eq. 2)}$$

When considering only the lower layer, a sensor node which has D neighbors needs to store one node key, D pair-wise keys, and one region key. Therefore, the storage requirement for this lower layer is $D + 2$.

The efficiency of storage requirement for the proposed protocol in this paper can be compared with that of LEAP which is a similar symmetric key based management structure. LEAP requires a space of $3D + 2 + L$, where L is key chain storage area for authentication containing a predefined constant value. Thus without L, the storage requirement comparison can be shown in graph as in figure 12. As the number of neighboring node increases the storage requirement for LEAP rises considerably while the proposed protocol only rises to one third of height.

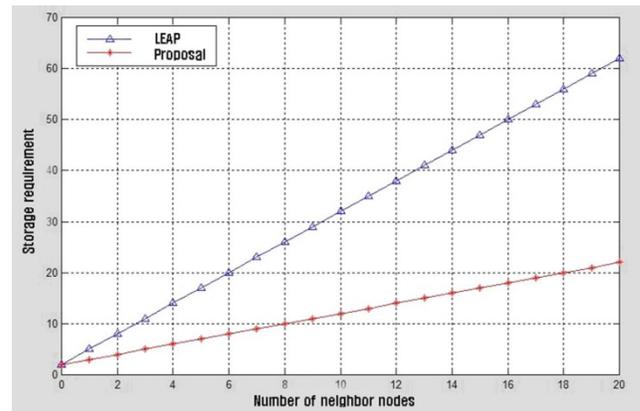

<Figure 12. Comparison of storage requirement>

This storage requirement is indeed an acceptable level considering products like Mica2 or MicaZ designed by Crossbow, one of the most popular sensor nodes, have 4KB of memory space available.

## VII. CONCLUSION AND FUTURE RESEARCH

This paper proposed a new efficient key management protocol for WSANs. The application of security is done in layers or hierarchical method. The upper layer which has less resource limitation is proposed to use security scheme based on the Public Key algorithm, while lower layer with high resource limitation uses scheme based on the Symmetric Key algorithm – namely pair-wise, node, and region keys.

Appropriate and efficient key structure for WSANs has been defined, detailed explanations of how each key is generated, and applied. Our method of key management can be applied to sensor actor networks.

Later researches will analyze more precisely the safety and effectiveness of this proposed method by using simulations. Also, distributed CA environment using methods such as Threshold Cryptography will be considered in case there is no sink node.

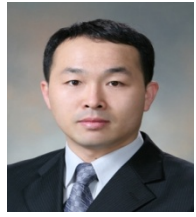

Yunho Lee received his B.S. in Electronic Engineering from Korea Military Academy in 1999, M.S. in Computer Engineering from Seoul National University, Korea, in 2005. Currently, he is a Ph.D. course in Computer Science, Korea National Defense University. His research interests include mobile network security, and intrusion detection.

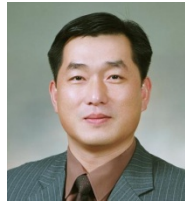

Soojin Lee received his B.S. in Computer Science from Korea Military Academy in 1992, M.S. in Computer Science from Younsei University, Korea, in 1996, and Ph.D. in Computer Science from Korea Advanced Institute of Science and Technology (KAIST), Korea in 2006. Since 2006, he has been an associate professor at the Dept. of Computer Science, Korea National Defense University. His research interest includes computer and communication security, intrusion detection, and mobile network security.